\documentclass[useAMS,usenatbib]{mn2e}
\usepackage{psfig, epsf, epsfig}

\title[Radio lobes and X-ray hot spots of an FRII microquasar]{Radio lobes and X-ray hot spots in the microquasar S26}
\author[R. Soria et al.]{Roberto Soria$^{1}$\thanks{E-mail:
roberto.soria@mssl.ucl.ac.uk},
Manfred W. Pakull$^{2}$, Jess W. Broderick$^{3}$, 
Stephane Corbel$^{4}$, 
\newauthor
Christian Motch$^{2}$\\
$^{1}$Mullard Space Science Laboratory, University College London, Holmbury St Mary, Surrey RH5 6NT, UK\\
$^{2}$University of Strasbourg, CNRS UMR 7550, Observatoire Astronomique, 
11 rue de l'Universit\'e, 67000 Strasbourg, France\\
$^{3}$School of Physics \& Astronomy, University of Southampton, Southampton, Hampshire SO17 1BJ, UK\\
$^{4}$Universit\'e Paris 7 and Service d'Astrophysique, UMR AIM, CEA Saclay, F-91191, Gif sur Yvette, France}

\begin{document}

\date{Accepted ... Received ... in original form ...}

\pagerange{\pageref{firstpage}--\pageref{lastpage}} \pubyear{2010}

\maketitle

\label{firstpage}

\begin{abstract}
We have studied the structure and energetics of the powerful 
microquasar/shock-ionized nebula S26 in NGC\,7793, with particular 
focus on its radio and X-ray properties. 
Using the Australia Telescope Compact Array, 
we have resolved for the first time the radio lobe structure 
and mapped the spectral index of the radio cocoon.  
The steep spectral index of the radio lobes is consistent with 
optically-thin synchrotron emission; outside the lobes, the spectral 
index is flatter, suggesting an additional contribution from free-free 
emission, and perhaps ongoing ejections near the core. 
The radio core is not detected, while the X-ray core has a $0.3$--$8$ keV 
luminosity $\approx 6 \times 10^{36}$ erg s$^{-1}$.
The size of the radio cocoon matches that seen in the optical 
emission lines and diffuse soft X-ray emission.
The total 5.5-GHz flux of cocoon and lobes is $\approx 2.1$ mJy, which 
at the assumed distance of 3.9 Mpc corresponds to about 3 times 
the luminosity of Cas A. The total 9.0-GHz flux is $\approx 1.6$ mJy.
The X-ray hot spots (combined $0.3$--$8$ keV luminosity 
$\approx 2 \times 10^{37}$ erg s$^{-1}$) are located $\approx 20$ pc 
outwards of the radio hot spots ({\it i.e.}, downstream along 
the jet direction), consistent with a different physical origin 
of X-ray and radio emission (thermal-plasma and synchrotron, respectively). 
The total particle energy in the bubble is $\sim 10^{53}$ erg: from the observed radio 
flux, we estimate that only $\sim$ a few $10^{50}$ erg are stored in the relativistic 
electrons; the rest is in protons, nuclei and non-relativistic electrons. 
The X-ray-emitting component 
of the gas in the hot spots contains $\sim 10^{51}$ erg, and 
$\sim 10^{52}$ erg over the whole cocoon. We suggest that S26 provides a clue 
to understand how the ambient medium is heated by the mechanical power 
of a black hole near its Eddington accretion rate.
\end{abstract}


\section{Introduction}

The basic physical model for radio lobes in FRII radio galaxies 
is based on a pair of relativistic, collimated jets emerging 
from the active black hole (BH). As the jet interacts with and is decelerated 
by the ambient (interstellar or intergalactic) medium, a reverse shock 
propagates inwards into the ejected plasma. After crossing the reverse 
shock, the jet material inflates a cocoon of hot gas, which is 
less dense but much overpressured with respect to the undisturbed 
medium. Thus, the cocoon expands supersonically, 
driving a forward shock (bow shock) into the ambient medium 
\citep{sch74,bla74,beg84,raw91,kai97}.   
The cocoon and lobes are the main sources of optically-thin 
(steep spectrum) synchrotron radio emission, while we expect 
optically-thick (flat-spectrum) radio emission from 
the jet near the core. A radio- and sometimes X-ray-luminous 
hot spot is usually found at the reverse shock, at the end 
of the jet. This is where 
most of the bulk kinetic energy of the jet is transferred 
to thermal ions, and 
to a non-thermal population of ultra-relativistic electrons, 
which cool via synchrotron and synchrotron self-Compton 
emission. 
Non-thermal X-ray emission at the hot spot position may be due 
to synchrotron and synchrotron self-Compton emission.
Optically-thin thermal plasma X-ray emission 
may come instead from the hot, shocked ambient gas  
between the reverse shock and the bow shock; 
in this case, the peak of the thermal 
X-ray emission will appear just in front of the radio hot spots. 

There is a scale invariance between the jet emission processes
in microquasars (powered by stellar-mass BHs)
and in AGN/quasars (powered by supermassive BHs). 
There is also at least one important difference: microquasars are
mostly located in a relatively low-pressure medium as compared
to the medium around AGN, when scaling of the jet
thrust is taken into account \citep{hei02}. 
As a consequence, we expect to see fewer, dimmer cocoons and radio lobes 
in microquasars than in the most powerful AGN and quasars;  
however, the linear sizes of those microquasar cocoons and jets 
can be up to 1000 times larger than in radio galaxies, 
scaled to their respective BH masses. There is also evidence that 
some microquasars are located inside low-density cavities, 
compared with the undisturbed interstellar medium \citep{hao09}.

So far, our knowledge of the interaction of microquasar jets
with the interstellar medium has largely relied on the
Galactic microquasar SS\,433 \citep{fab04} and its surrounding 
synchrotron-emitting nebula W50 (size $\sim 100 \times 50$ pc).
A mildly relativistic ($v_J = 0.27c$), precessing jet acts
as a sprinkler that inflates ``ear-like'' lobe structures,
protruding from the more spherical W50 nebula. Most of the jet
power ($\sim 10^{39}$ erg s$^{-1}$)
is dissipated in the lobes \citep{beg80}.
Faint evidence of the interaction of relativistic jets
with the interstellar medium has been found in a few
other, less powerful Galactic microquasars: for example 
Cyg X-1 \citep{gal05}, 
GRS\,1915$+$105 \citep{kai04}, XTE\,J1550$-$564 \citep{cor02}, 
H1743$-$322 \citep{cor05}, and around the neutron star Sco X-1 
\citep{fom01}.
On a larger scale, huge (size $\ga 100$ pc) ionized nebulae 
have been found around several ultraluminous
X-ray sources (ULXs) in nearby galaxies \citep{pak02,pak03,rob03,pak06,pak08,gri08,fen08}. 
Such nebulae emit optical lines
typical of shock-ionized gas, and in a few cases,
synchrotron radio emission \citep{mil05,sor06,lan07}. The derived ages
($\ga$ a few $10^5$ yr) and energy content ($\sim 10^{52}$--$10^{53}$ erg)
are too large for ordinary supernova remnants, and suggest jet/wind inflation
with a mechanical power $\sim 10^{39}$--$10^{40}$ erg s$^{-1}$,
comparable with the X-ray luminosities \citep{pak06}.
However, no direct X-ray or radio evidence of a collimated jet 
has been found in ULX bubbles so far.
On the other hand, X-ray luminous sources may be only
a subset of non-nuclear BHs at very high mass accretion rates.
\citet{pak08} proposed that ionized bubbles might also be found
associated with BHs that appear X-ray faint, either because their
radiative emission is collimated away from our line of sight,
or because they are transients and currently in a low/off accretion
state, or because they channel most of their accretion power into a jet 
even at near-Eddington mass accretion rates.

\begin{figure}
\begin{center}
\psfig{figure=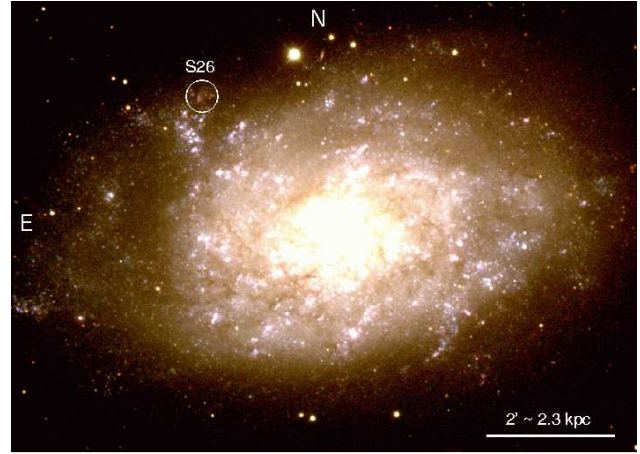,width=83mm,angle=0}
\end{center}
\caption{Location of S26 inside its host galaxy NGC\,7793, 
from public-archive $BVR$ images taken from the Cerro Tololo Inter-American 
Observatory (CTIO) 1.5-m telescope on 2001 October 18. 
S26 is inside the circle (radius of $15\arcsec$) 
in the north-eastern corner of the galaxy.}
\label{f2}
\end{figure}

\begin{figure}
\begin{center}
\psfig{figure=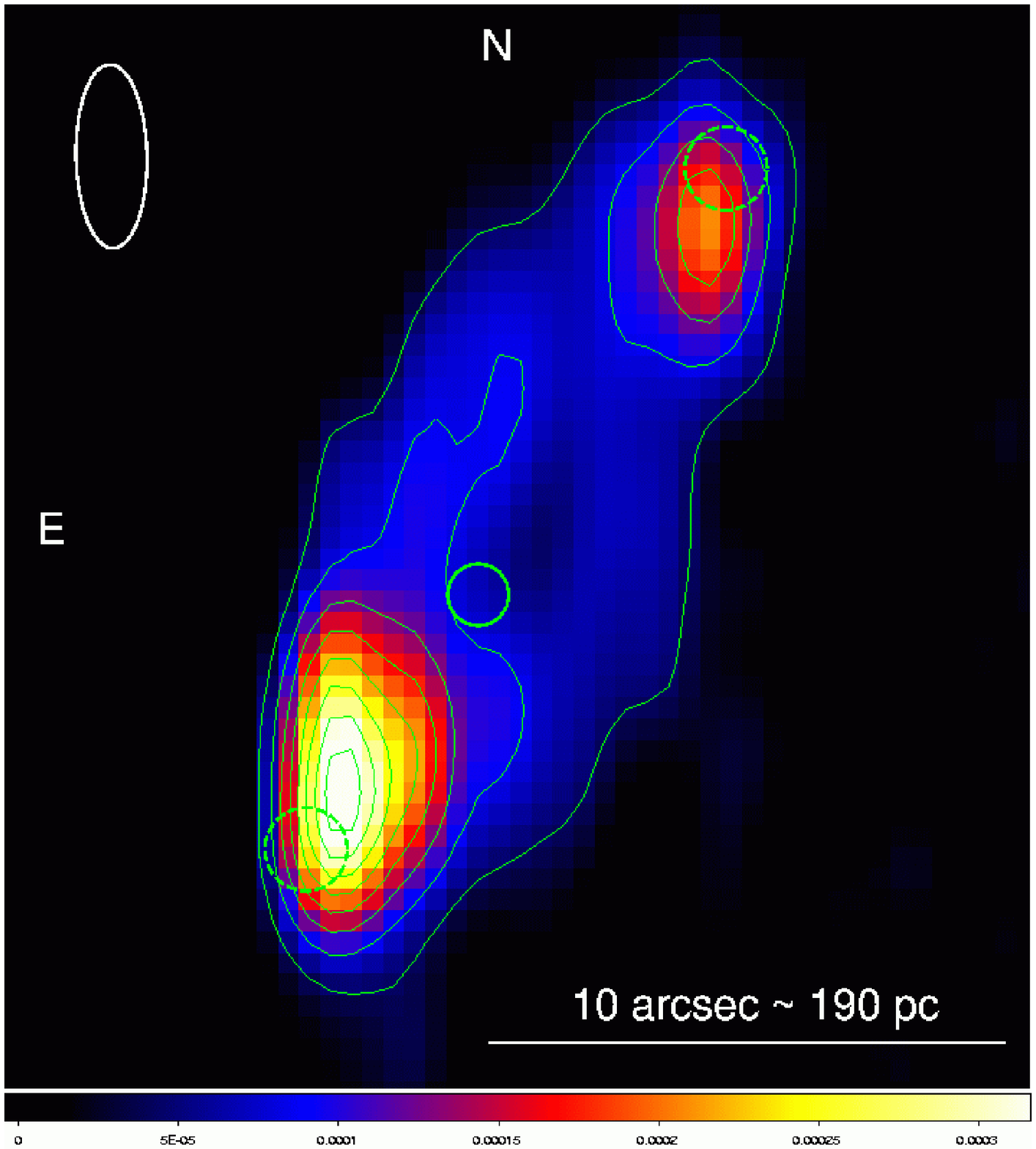,width=83mm,angle=0}
\psfig{figure=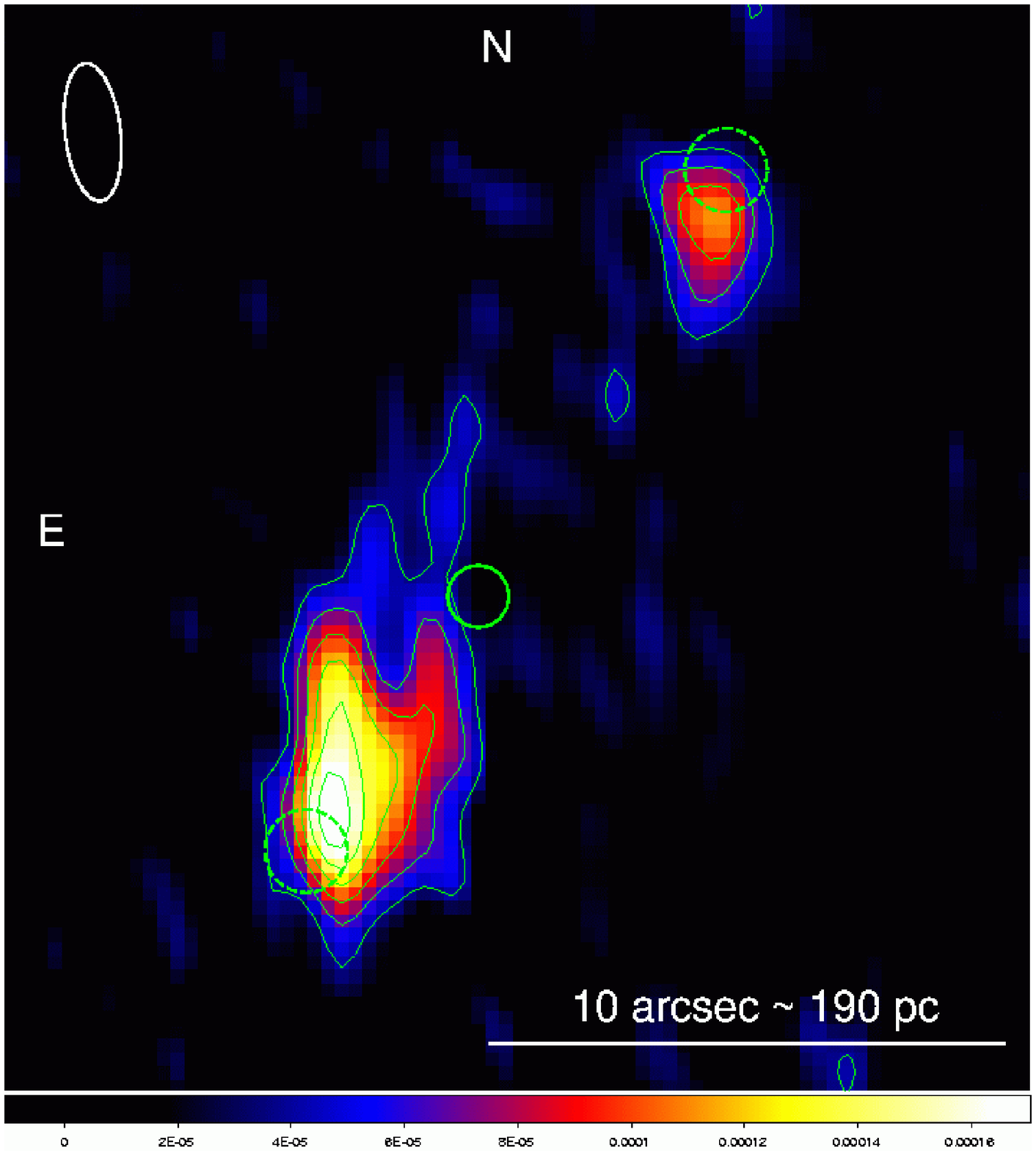,width=83mm,angle=0}
\end{center}
\caption{Top panel: ATCA map at 5.48 GHz, with intensity scale 
   in Jy beam$^{-1}$. Contour intensity levels are ($0.40,0.84,1.29,1.73,2.17,2.61,3.06,3.50$)
   $\times 10^{-4}$ Jy beam$^{-1}$. The rms noise level is $0.085 \times 10^{-4}$ 
   Jy beam$^{-1}$. Overplotted green circles mark the positions of the X-ray core 
   and hot spots ($0\farcs6$ radius for the core, $0\farcs8$ for the hot spots), 
   from {\it Chandra}.
   Bottom panel:  ATCA map at 9.02 GHz, with intensity scale 
   in Jy beam$^{-1}$. Contour intensity levels are  
   ($0.39,0.65,0.91,1.17,1.44,1.70$)
   $\times 10^{-4}$ Jy beam$^{-1}$. The rms noise level is $0.135 \times 10^{-4}$ 
   Jy beam$^{-1}$.}
\label{f1}
\end{figure}

\section{The microquasar S26 in NGC\,7793}

A spectacular example of such systems was recently discovered 
\citep[][henceforth PSM10]{pak10}
in the outskirts of the Sculptor galaxy NGC\,7793 (Figure 1), 
at a distance of 3.9 Mpc \citep{kar03}.
The radio/optical nebula S26 was originally classified as
a supernova remnant candidate \citep{bla97}; the high
[S\,{\footnotesize{II}}] $\lambda 6716,6732/{\rm H}\alpha$ flux 
ratio indicates the presence of shock-ionized gas. 
The optical radial velocity of S26 agrees with that of NGC\,7793, 
ruling out a chance superposition of a background AGN.
A radio spectral index consistent with optically-thin synchrotron emission 
was reported by \citet{pan02}, and the emitting region appeared 
clearly extended and elongated. However, the spatial resolution 
was too low to reveal details of its internal structure.
A faint X-ray source was discovered to be associated with S26 in 
{\it ROSAT} observations \citep{rea99}, but it was unresolved.
Using {\it Chandra} data, \citet{pak08} discovered
that the X-ray emission is resolved into three sources 
that are perfectly aligned and match the extent of the major axis 
of the radio and optical nebulae. Those sources have been interpreted 
as the core (at the X-ray binary position) and the X-ray hot spots (where 
the jet interacts with the ambient medium). 

From optical spectroscopic observations, PSM10 determined 
the expansion velocity, density and temperature of the 
line-emitting gas in the bubble, and discovered that the mechanical power 
of the central BH is $\sim$ a few $\times 10^{40}$ erg s$^{-1}$; 
this suggests accretion rates similar to those required for 
the most luminous ULXs. PSM10 showed that the jet power 
is orders of magnitude higher than both the X-ray luminosity and  
the value one would derive from the radio luminosity; they argued 
that most of the jet power is transferred to non-relativistic 
protons and nuclei rather than non-thermal relativistic electrons.

In this paper, we present the initial results of our radio study, 
showing for the first time the resolved lobe structure and measuring 
the spectral index variations across the source. 
We discuss the origin of the radio emission 
and the implied jet power. We compare the radio, X-ray and optical 
maps of the nebula, determining the positions of the radio and X-ray 
hot spots and of the core, and we provide a more detailed spectral analysis 
and interpretation of the X-ray properties. We then summarize the energy 
budget of this system, quantifying the fraction of energy 
stored in relativistic electrons and in the X-ray emitting gas.

\begin{figure}
\begin{center}
\psfig{figure=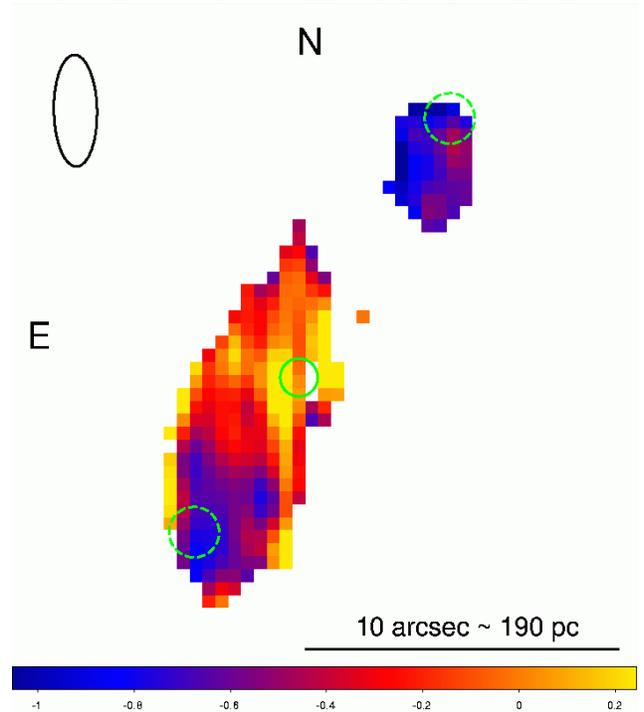,width=83mm,angle=0}
\end{center}
\caption{Map of the radio spectral index, inferred from the ratio of the 5.48 GHz 
and 9.02 GHz maps, where the latter has been tapered to the resolution of the former. 
Green circles mark the positions of the X-ray core and X-ray hot spots.}
\label{f2}
\end{figure}

\begin{figure}
\begin{center}
\psfig{figure=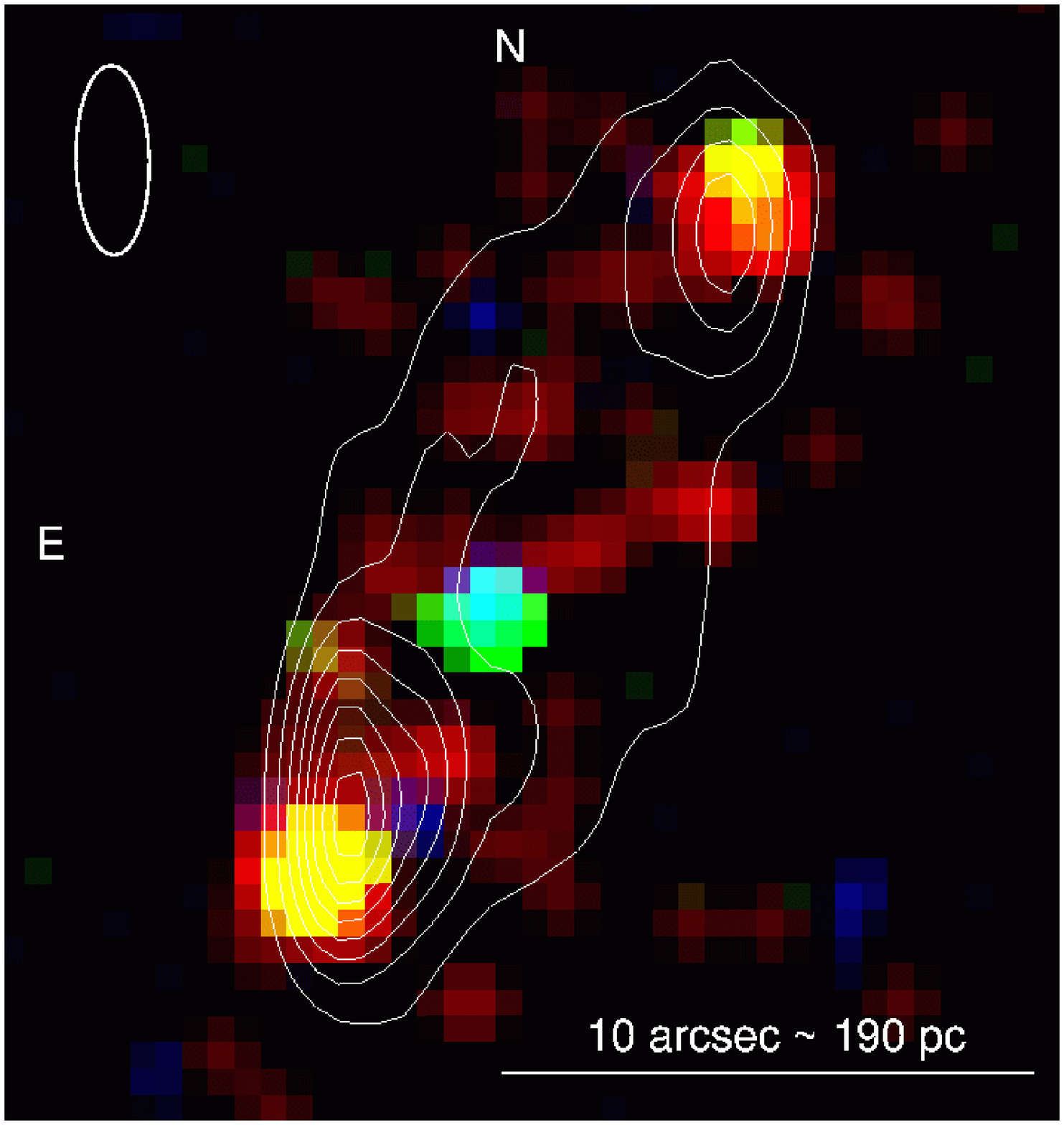,width=83mm,angle=0}
\psfig{figure=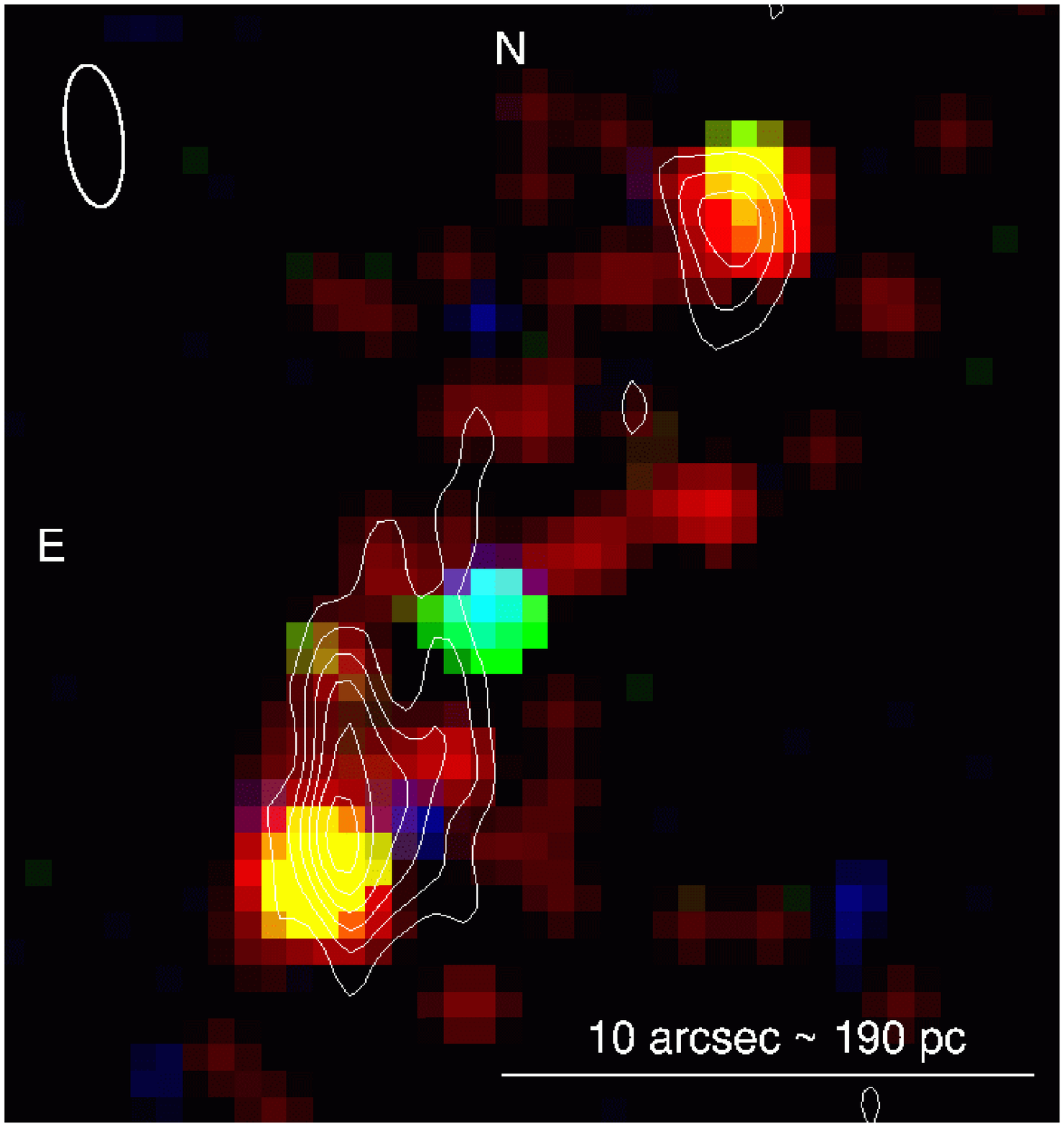,width=83mm,angle=0}
\end{center}
\caption{Top panel: {\it Chandra}/ACIS color map of S26
(smoothed with a $1\arcsec$ Gaussian core), with ATCA 5.48-GHz 
intensity contours. The colour coding is: red: $0.2$--$1$ keV; green: $1$--$2$ keV; blue: $2$--$8$ keV.
Contour intensity levels are as in Figure 2.
Bottom panel: {\it Chandra}/ACIS X-ray colour map (same colour coding), with 
ATCA 9.02-GHz radio intensity contours superimposed. Contour intensity levels are as in Figure 2.
Cf.~Fig.~1 in PSM10, which overplots H$\alpha$ emission contours over the same {\it Chandra} image.} 
\label{f3}
\end{figure}

\section{Observations}

\subsection{Radio observations}

We observed S26 on 2009 August 6 and 7 with the Australia Telescope Compact Array (ATCA). 
Simultaneous 5.5 and 9 GHz observations were carried out with 
the Compact Array Broadband Backend (CABB); the bandwidth at each frequency 
is about 2 GHz. The array configuration was 6D, with minimum and maximum 
baselines of 77 and 5878 m, respectively. The total integration time 
on-source was 13.3 hours; the data for antenna 6 at 9 GHz were lost 
during part of the second observing session due to technical problems.
B1934$-$638 was used as the primary calibrator, 
while our secondary calibrator was B2357$-$318.

We reduced and imaged the data with {\small MIRIAD} \citep*[][]{sault95}. 
After flagging bad data, the effective frequencies of the two bands are 5.48 and 9.02 GHz. 
We tried different values of Briggs' robust weighting parameter 
\citep[][]{briggs95}; we found that a value of 0.0 provides a good balance 
between sidelobe suppression and sensitivity 
at both frequencies. Because of the wide bandwidths, we used the multi-frequency 
deconvolution algorithm {\small MFCLEAN} \citep[][]{sault94}. The {\small CLEAN}ed, 
primary-beam-corrected images are shown in Figure 2; the angular resolutions 
are $3\farcs54 \times 1\farcs38$ (position angle $1.1^{\circ}$) 
and $2\farcs67 \times 1.08$ (position angle $6.8^{\circ}$) 
at 5.48 and 9.02 GHz, respectively. In the vicinity of the microquasar, 
the rms noise levels are $8.5 \mu$Jy beam$^{-1}$ (5.48 GHz) 
and $13.5 \mu$Jy beam$^{-1}$ (9.02 GHz). We estimate that the internal 
calibration uncertainty is $\sim 2$ per cent at both frequencies.

We also tapered the 9.02 GHz data so that the resolution and beam position 
angle matched those of the 5.48 GHz data. 
We created a two-point spectral index map (Figure 3), 
where the sign of the index is defined such that 
the specific flux $S_{\nu} \sim \nu^{\alpha}$.
For the tapered 9.02 GHz data, we found that a robust weighting parameter 
of 0.5 provides the best compromise between residual sidelobe contamination and  
sensitivity to the low-surface-brightness extended emission that is 
clearly visible in the 5.5 GHz map.


\subsection{X-ray observations}

NGC\,7793 was observed with {\it Chandra}/ACIS-S3 on 
2003 September 6 (Obs ID 3954). The live time was 48.9 ks.
We retrieved the data from the public archives (processed 
with ASCDVER=7.6.8), and analysed them with standard imaging 
and spectroscopic tools such as {\it psextract} in 
the data analysis system {\footnotesize CIAO} Version 4.0 
\citep{2006SPIE.6270E..60F}. We modelled the X-ray spectra 
with {\footnotesize XSPEC} Version 12.0 \citep{arn96}.
Luckily, the roll angle of the {\it Chandra} observation was such 
that S26 was located rather close to the S3 aimpoint 
(less than $1\arcmin$ away), giving us a narrower 
point spread function.


\section{Main results}

\subsection{Radio results}

The most important new result of our ATCA study 
is that we have resolved the spatial structure of the radio-emitting nebula.
Most of the emission comes from two radio hot spots and surrounding lobes, 
with a fainter but clearly identified cocoon encompassing them (Figure 2, 
top panel). This is the textbook structure \citep[{\it {e.g.}},][]{beg84} 
of FR\,II-type powerful radio galaxies \citep[{\it {e.g.}}, Cygnus A:][]{car96,wsy06}.
The radio structure is aligned with the jet axis suggested by the 
three X-ray sources, confirming this interpretation. 
The position of the northern radio hot spot is 
RA $= 23^h 57^m 59^s.58$, Dec $=-32^{\circ} 33' 13\farcs6$
(with an uncertainty of $\approx 0\farcs2$).
The position of the southern hot spot is 
RA $= 23^h 58^m 00^s.15$, Dec $=-32^{\circ} 33' 25\farcs0$.
Thus, the projected distance between the radio hot spots  
is $(13\farcs5 \pm 0\farcs3) \approx 250$ pc. 
We interpret the radio hot spots as the reverse shocks 
(Mach disks) at the ends of the jets.

At 5.5 GHz, the peak intensity in the southern lobe 
is $\approx 0.37$ mJy beam$^{-1}$; in the northern lobe, $\approx 0.21$ mJy beam$^{-1}$;
the total flux in the lobes and cocoon is $\approx 2.1$ mJy (Table 1), that is $\approx 3$ times 
the luminosity of Cas A. 
From the untapered map at 9 GHz, we obtain a peak intensity in the southern lobe 
$\approx 0.19$ mJy beam$^{-1}$; in the northern lobe, $\approx 0.11$ mJy beam$^{-1}$.
The total flux at 9 GHz is $\approx 1.6$ mJy (Table 1).
The spectral index in the lobes is, on average, $\approx -0.7$ to $-0.6$;
it appears to be flatter ($\approx -0.4$ to $0$) across most of the cocoon, 
and inverted ($\approx 0$ to $0.4$) at the base of the jets, on either side 
of the X-ray/optical core (Figure 3).
We estimate a $1\sigma$ uncertainty for $\alpha$ of $\approx 0.12$ near the southern 
radio hot spot, $\approx 0.19$ near the northern radio hot spot, 
and $\approx 0.5$--$0.6$ in the rest of the cocoon, where the emission 
is much fainter. Thus, the existence of a complex spatial structure 
for the spectral index is at this stage still an intriguing speculation 
that has to be tested with deeper observations.

\begin{figure*}
\begin{center}
\psfig{figure=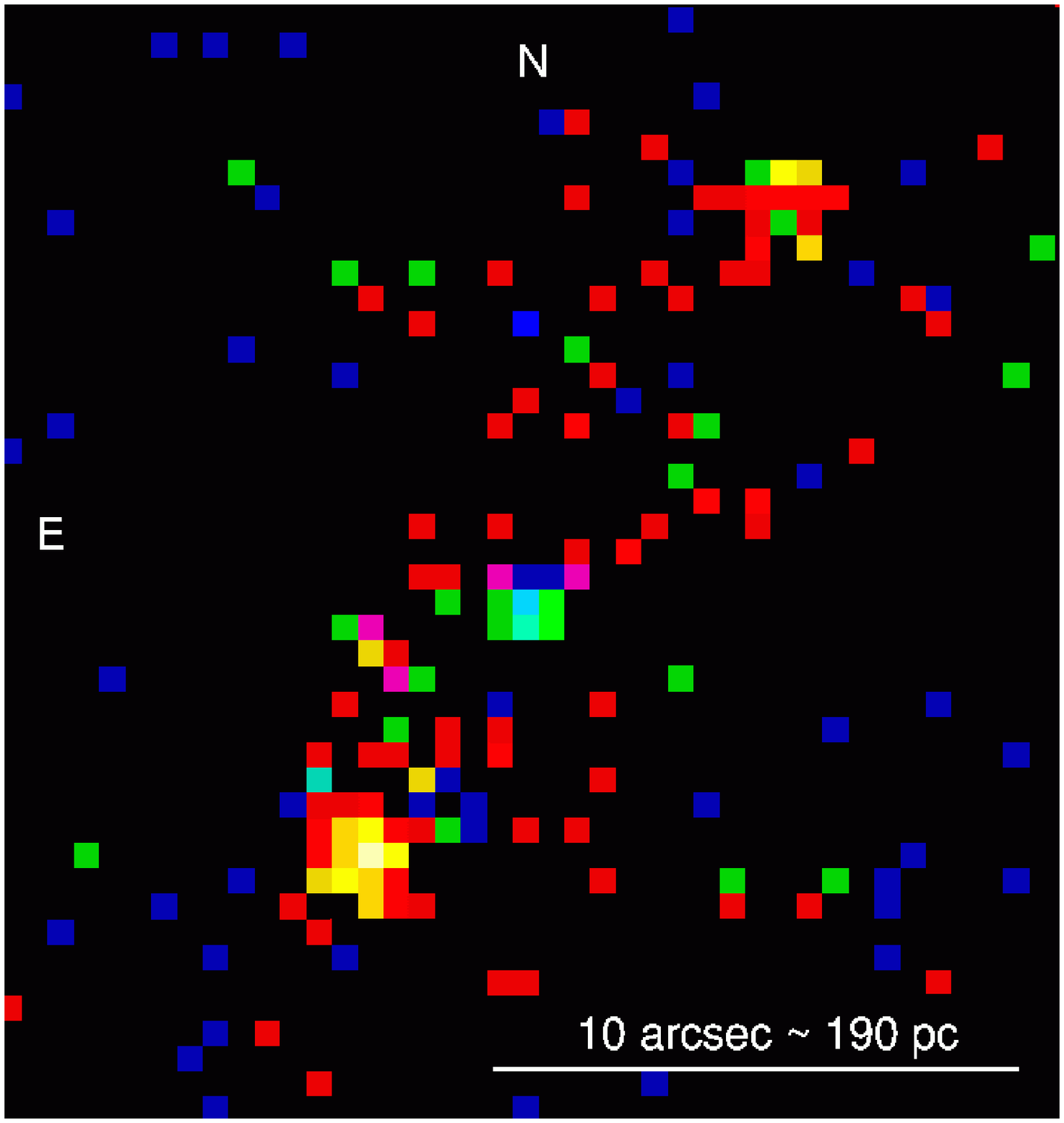,width=58mm,angle=0}
\psfig{figure=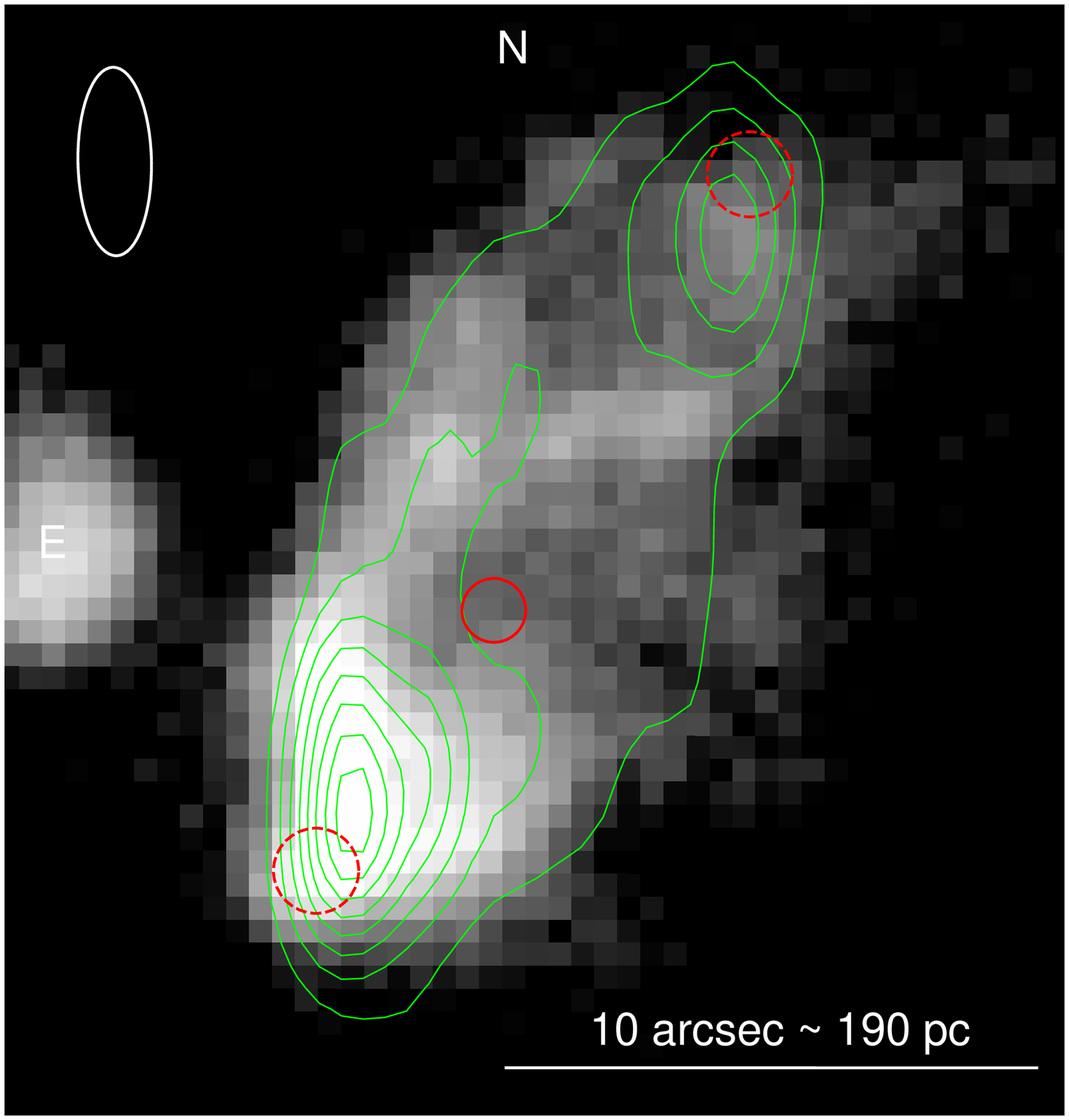,width=58mm,angle=0}
\psfig{figure=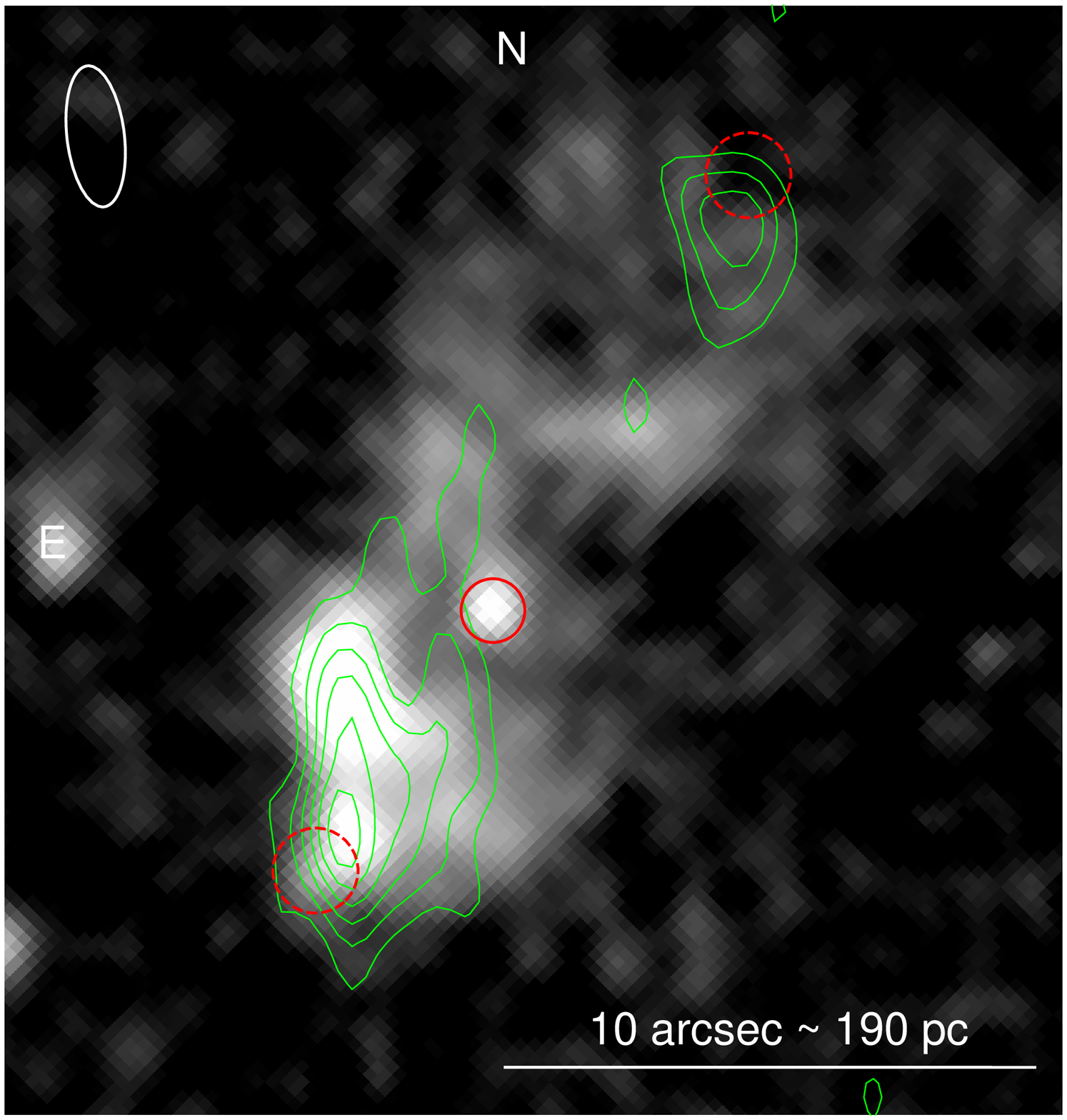,width=58mm,angle=0}
\end{center}
\caption{Left panel: unsmoothed {\it Chandra} color map, with the same colour 
coding as in Figure 4. 
Middle panel: greyscale, continuum-subtracted H$\alpha$ emission, 
with ATCA 5.5-GHz intensity contours. Red circles mark the positions of the X-ray core 
and hot spots ($0\farcs6$ radius for the core, $0\farcs8$ for the hot spots). 
The public-archive H$\alpha$ image was taken from the CTIO 
1.5-m telescope on 2001 October 18.
Right panel: greyscale, continuum-subtracted HeII $\lambda 4686$ emission, 
with ATCA 9.0-GHz intensity contours and X-ray core/hot spot positions. 
The image was taken with the FORS1 camera on the ESO Very Large Telescope 
on 2002 November 1; we smoothed it with a $0\farcs6$ Gaussian core. See PSM10 
for a discussion of the bright optical core.} 
\label{f4}
\end{figure*}

\begin{table}
\begin{center}
\begin{tabular}{lrr}\hline
Structure & $S_{5.5}$ (mJy) & $S_{9.0}$ (mJy)\\
\hline
Total & $2.1 \pm 0.1$ & $1.6 \pm 0.1$ \\[3pt]
S lobe & $0.95 \pm 0.05$ & $0.80 \pm 0.05$ \\[3pt]
N lobe & $0.45 \pm 0.03$ & $0.28 \pm 0.03$ \\[3pt]
\hline
\end{tabular}
\end{center}
\caption{Observed radio fluxes for the whole nebula, 
and for the two lobes, from our ATCA observations. 
The 9.0-GHz fluxes are measured from the tapered map.}
\label{tab:a}
\end{table}

\subsection{X-ray results}

The key feature of this system is the aligned triplet 
of point-like sources (Figures 4, 5), which we interpret as the X-ray core 
and hot spots \citep{pak08}. The X-ray core is located at 
RA $= 23^h 57^m 59^s.94$, Dec $=-32^{\circ} 33' 20\farcs9$
(with an uncertainty of $\approx 0\farcs2$).
It has a hard spectrum (power-law photon index $\Gamma = 1.4 \pm 0.6$), 
consistent with a BH in the low/hard state \citep{rem06}, 
and an emitted luminosity $L_{\rm 0.3-8} \approx 6 \times 10^{36}$
erg s$^{-1}$ (Table 2 and Figure 6).
The X-ray core coincides, within the astrometric uncertainties, 
with a point-like optical source 
with HeII $\lambda 4686$ emission (Figure 5). 
The 90\% uncertainty circle 
of the ACIS-S3 absolute position has a radius 
of $0\farcs4$\footnote{http://cxc.harvard.edu/cal/ASPECT/celmon/}, 
and the uncertainty of the optical images is $\approx 0\farcs3$.

The two hot spots have a much softer spectrum (Figures 4, 6), 
and are well fitted (Table 3)
by a 2-component {\it raymond-smith} thermal plasma model \citep{ray77} with
$kT_1 \approx 0.3$ keV and $kT_2 \approx 0.9$ keV, 
and negligible intrinsic absorption 
(Cash statistics $=10.4$ over 13 dof 
for solar abundances; $=9.6$ over 13 dof 
for 1/4-solar abundances). 
The emitted X-ray luminosities 
are $L_{\rm 0.3-8} \approx 5 \times 10^{36}$
erg s$^{-1}$ and $L_{\rm 0.3-8} \approx 11 \times 10^{36}$
erg s$^{-1}$ for the northern and southern hot spot, respectively 
(similar to the ratio of radio luminosities).
In general, a choice of low metal abundances give better fits 
than solar abundances, but the signal-to-noise is not high enough 
to contrain this parameter.
Other, more complex thermal plasma models such as {\it mekal}, {\it vmekal}, 
{\it equil} and {\it nei} also give similar sets of best-fitting 
parameters; they all require at least two temperature components.
However, the {\it sedov} thermal plasma model \citep{bor01} 
gives a good fit (Cash statistics $=11.9$ over 15 dof 
for 1/4-solar abundances) with only one temperature component, at
$kT \approx 0.52$ keV (Table 4); the ionization age ($\tau$ parameter) 
of the {\it sedov} model is consistent with the characteristic age 
multiplied by electron density in the S26 bubble.
Given the small number of counts in the {\it Chandra} spectrum, 
none of the more complex thermal-plasma models can provide an improvement 
over the simpler {\it raymond-smith} model.
Simple or broken power-law models do not give acceptable fits 
(Cash statistics $=25.5$ over 15 dof); 
moreover, they would require an unphysically steep slope 
($\Gamma \approx 6$) combined with high intrinsic column densities 
($N_H \approx 5 \times 10^{21}$ cm$^{-2}$). We conclude that 
the hot spot spectra are not dominated by synchrotron or synchrotron 
self-Compton emission. We interpret them as optically-thin 
thermal plasma emission from hot, shocked gas, probably located between the reverse 
and forward shocks. The radius of the X-ray hot spots is $\la 1\arcsec$.
From the combined volume of the two hot spots and their emission 
measures (Table 3), we estimate  
a hot gas density $\ga 1$ cm$^{-3}$ and a mass $\sim 10^{36}$ g 
(see also PSM10).

We also find faint X-ray emission projected over the surface 
of the cocoon (Figure 5), with slightly softer colours 
than the hot spots; however, the number of detected counts is too low 
for detailed temperature comparisons. A single-temperature bremsstrahlung 
fit suggests $kT = 0.5 \pm 0.1$ keV (Table 5 and Figure 7). It is also impossible to determine 
at this stage whether the X-ray emitting gas is filling the whole cocoon, 
or is confined to an outer shell. If the hot gas fills a significant fraction 
of the cocoon, its characteristic density 
is $\sim 0.1$ cm$^{-3}$ and the total mass $\sim 10^{37}$ g.

The position of the northern X-ray spot is 
RA $= 23^h 57^m 59^s.56$, Dec $=-32^{\circ} 33' 12\farcs6$
(with an uncertainty of $\approx 0\farcs2$).
The position of the southern X-ray spot is 
RA $= 23^h 58^m 00^s.20$, Dec $=-32^{\circ} 33' 25\farcs7$.
Therefore, the projected distance between the X-ray hot spots 
is $(15\farcs4 \pm 0\farcs3) \approx 290$ pc, 
slightly larger than the projected distance between the radio hot spots; 
that is, each X-ray hot spot appears projected $\approx 15$--$20$ pc 
downstream from the corresponding radio hot spot.
In fact, the X-ray and radio hot spot positions in the southern lobe 
coincide with two distinct brightness peaks in the optical images, 
particularly H$\alpha$ and $V$ band (Figures 5, 8).
This provides another argument in favor of two separate 
emission mechanisms for the radio and X-ray hot spots.
The core is along the line of the hot spots, but not symmetrically 
located between them. It is projected $\approx 6\arcsec$ from the southern spot 
(which is about twice as luminous in every band) and $\approx 9\farcs5$ from 
the northern spot. This may perhaps be due to a higher density of the 
interstellar medium south of the core.

\begin{table}
\begin{center}
\begin{tabular}{lr}
\hline
Parameter & Value\\
\hline
$N_{H,\rm Gal}$ & \ \ \ \ $1.2 \times 10^{20}$ cm$^{-2}$ \ \ (fixed)\\[3pt]
$\Gamma$ & \ \ \ \ $1.4^{+0.6}_{-0.6}$ \\[3pt]
$K$ & \ \ \ \ $\left(4.3^{+2.1}_{-1.6}\right) \times 10^{-7}$\\[3pt]
\hline
$f_{0.3-8}$ & \ \ \ \ $\left(3.4^{+0.8}_{-0.8}\right) \times 10^{-15}$ erg cm$^{-2}$ s$^{-1}$\\[3pt]
$L_{0.3-8}$ & \ \ \ \ $\left(6.2^{+1.5}_{-1.5}\right) \times 10^{36}$ erg s$^{-1}$\\[3pt]
\hline
C-statistic &  \ \ \ \  $4.56$ (6 dof)\\[3pt]
\hline
\end{tabular}
\end{center}
\caption{Best-fitting spectral parameters for the X-ray emission 
from the core. The XSPEC model is {\it wabs$_{\rm Gal}$*powerlaw}. 
Adding intrinsic absorption does not improve the fit.
Errors are 90\% confidence level for 1 interesting parameter. 
Here, and in the following tables, the line-of-sight column density 
to NGC\,7793 is from \citet{kal05}.}
\label{tab:a}
\end{table}

\begin{table}
\begin{center}
\begin{tabular}{lr}\hline
Parameter & Value\\
\hline
$N_{H,\rm Gal}$ & \ \ \ \ $1.2 \times 10^{20}$ cm$^{-2}$ \ \ (fixed)\\[3pt]
$N_H$ & \ \ \ \ $< 1.0 \times 10^{21}$ cm$^{-2}$ \\[3pt]
$Z$  & \ \ \ \  1  \ \ (fixed)\\[3pt]
$kT_1$ & \ \ \ \ $0.26^{+0.05}_{-0.08}$ keV\\[3pt]
$N_1$ & \ \ \ \ $\left(2.3^{+2.9}_{-0.6}\right) \times 10^{-6}$\\[3pt]
$kT_2$ & \ \ \ \ $0.96^{+0.31}_{-0.17}$ keV\\[3pt]
$N_2$ & \ \ \ \ $\left(1.9^{+0.7}_{-0.7}\right) \times 10^{-6}$\\[3pt]
\hline
$f_{0.3-8}$ & \ \ \ \ $\left(8.8^{+0.9}_{-0.9}\right) \times 10^{-15}$ erg cm$^{-2}$ s$^{-1}$\\[3pt]
$L_{0.3-8}$ & \ \ \ \ $\left(1.7^{+1.2}_{-0.3}\right) \times 10^{37}$ erg s$^{-1}$\\[3pt]
EM(0.26 keV) & $\left(4.2^{+5.3}_{-1.1}\right) \times 10^{59}$ cm$^{-3}$\\[3pt]
EM(0.96 keV) & $\left(3.5^{+1.3}_{-1.3}\right) \times 10^{59}$ cm$^{-3}$\\[3pt]
\hline
C-statistic &  \ \ \ \  $10.35$ (13 dof)\\[3pt]
\hline
\hline
$N_{H,\rm Gal}$ & \ \ \ \ $1.2 \times 10^{20}$ cm$^{-2}$ \ \ (fixed)\\[3pt]
$N_H$ & \ \ \ \ $< 1.1 \times 10^{21}$ cm$^{-2}$ \\[3pt]
$Z$  & \ \ \ \  0.25  \ \ (fixed)\\[3pt]
$kT_1$ & \ \ \ \ $0.29^{+0.08}_{-0.11}$ keV\\[3pt]
$N_1$ & \ \ \ \ $\left(7.3^{+12.2}_{-3.1}\right) \times 10^{-6}$\\[3pt]
$kT_2$ & \ \ \ \ $0.90^{+0.32}_{-0.16}$ keV\\[3pt]
$N_2$ & \ \ \ \ $\left(4.9^{+9.2}_{-1.9}\right) \times 10^{-6}$\\[3pt]
\hline
$f_{0.3-8}$ & \ \ \ \ $\left(9.1^{+0.9}_{-0.9}\right) \times 10^{-15}$ erg cm$^{-2}$ s$^{-1}$\\[3pt]
$L_{0.3-8}$ & \ \ \ \ $\left(1.8^{+1.2}_{-0.3}\right) \times 10^{37}$ erg s$^{-1}$\\[3pt]
EM(0.29 keV) & $\left(13.3^{+22.2}_{-5.6}\right) \times 10^{59}$ cm$^{-3}$\\[3pt]
EM(0.90 keV) & $\left(8.9^{+16.7}_{-3.5}\right) \times 10^{59}$ cm$^{-3}$\\[3pt]
\hline
C-statistic &  \ \ \ \  $9.55$ (13 dof)\\[3pt]
\hline
\end{tabular}
\end{center}
\caption{Best-fitting spectral parameters for the (combined) 
hot spot X-ray emission, assuming solar and 1/4-solar abundances. The XSPEC model 
is {\it wabs$_{\rm Gal}$*wabs*(ray+ray)}. Errors are 90\% confidence level 
for 1 interesting parameter.}
\label{tab:a}
\end{table}

\begin{table}
\begin{center}
\begin{tabular}{lr}\hline
Parameter & Value\\
\hline
$N_{H,\rm Gal}$ & \ \ \ \ $1.2 \times 10^{20}$ cm$^{-2}$ \ \ (fixed)\\[3pt]
$N_H$ & \ \ \ \ $< 2.5 \times 10^{21}$ cm$^{-2}$ \\[3pt]
$Z$  & \ \ \ \  0.25  \ \ (fixed)\\[3pt]
$kT$ & \ \ \ \ $0.78^{+0.07}_{-0.16}$ keV\\[3pt]
$N$ & \ \ \ \ $\left(8.4^{+8.7}_{-1.4}\right) \times 10^{-6}$\\[3pt]
\hline
C-statistic &  \ \ \ \  $28.9$ (15 dof)\\[3pt]
\hline
\hline
$N_{H,\rm Gal}$ & \ \ \ \ $1.2 \times 10^{20}$ cm$^{-2}$ \ \ (fixed)\\[3pt]
$N_H$ & \ \ \ \ 0  \ \ (fixed) \\[3pt]
$Z$  & \ \ \ \  0.25  \ \ (fixed)\\[3pt]
$kT_a$ & \ \ \ \ $0.52^{+0.19}_{-0.13}$ keV\\[3pt]
$kT_b$ & \ \ \ \ $kT_a$   \ \ (fixed)\\[3pt]
$\tau$ & \ \ \ \ $\left(1.9^{+11.1}_{-1.4}\right) \times 10^{12}$ s cm$^{-3}$\\[3pt]
$N_{\rm sed}$ & \ \ \ \ $\left(8.7^{+2.8}_{-2.8}\right) \times 10^{-6}$\\[3pt]
\hline
$f_{0.3-8}$ & \ \ \ \ $\left(9.0^{+1.6}_{-1.8}\right) \times 10^{-15}$ erg cm$^{-2}$ s$^{-1}$\\[3pt]
$L_{0.3-8}$ & \ \ \ \ $\left(1.8^{+0.3}_{-0.3}\right) \times 10^{37}$ erg s$^{-1}$\\[3pt]
EM & $\left(15.9^{+5.1}_{-5.1}\right) \times 10^{59}$ cm$^{-3}$\\[3pt]
\hline
C-statistic &  \ \ \ \  $11.88$ (15 dof)\\[3pt]
\hline
\end{tabular}
\end{center}
\caption{Alternative spectral models for the combined 
hot spot X-ray emission. A single-temperature {\it wabs$_{\rm Gal}$*wabs*ray} 
model does not produce acceptable fits. However, a single-temperature 
{\it wabs$_{\rm Gal}$*wabs*sedov} model results in a fit as good as 
those with two-temperature {\it raymond-smith} models (Table 3).
Errors are 90\% confidence level for 1 interesting parameter.}
\label{tab:a}
\end{table}



\begin{table}
\begin{center}
\begin{tabular}{lr}
\hline
Parameter & Value\\
\hline
$N_{H,\rm Gal}$ & \ \ \ \ $1.2 \times 10^{20}$ cm$^{-2}$ \ \ (fixed)\\[3pt]
$Z$  & \ \ \ \  0.25 \ \  (fixed)\\[3pt]
$kT_1$ & \ \ \ \ $0.29$ keV \ \  (fixed)\\[3pt]
$N_1$ & \ \ \ \ $\left(3.5^{+2.1}_{-1.8}\right) \times 10^{-6}$\\[3pt]
$kT_2$ & \ \ \ \ $0.90$ keV \ \  (fixed)\\[3pt]
$N_2$ & \ \ \ \ $\left(2.1^{+1.6}_{-1.3}\right) \times 10^{-6}$\\[3pt]
\hline
$f_{0.3-8}$ & \ \ \ \ $\left(4.2^{+0.6}_{-0.6}\right) \times 10^{-15}$ erg cm$^{-2}$ s$^{-1}$\\[3pt]
$L_{0.3-8}$ & \ \ \ \ $\left(7.6^{+1.1}_{-1.1}\right) \times 10^{36}$ erg s$^{-1}$\\[3pt]
EM(0.26 keV) & $\left(6.4^{+3.8}_{-3.3}\right) \times 10^{59}$ cm$^{-3}$\\[3pt]
EM(0.96 keV) & $\left(3.8^{+2.9}_{-2.4}\right) \times 10^{59}$ cm$^{-3}$\\[3pt]
\hline
C-statistic &  \ \ \ \  $11.28$ (17 dof)\\[3pt]
\hline
\hline
$N_{H,\rm Gal}$ & \ \ \ \ $1.2 \times 10^{20}$ cm$^{-2}$ \ \ (fixed)\\[3pt]
$kT_{\rm br}$ & \ \ \ \ $0.47^{+0.15}_{-0.13}$ keV \\[3pt]
$K_{\rm br}$ & \ \ \ \ $\left(7.5^{+5.9}_{-3.2}\right) \times 10^{-6}$\\[3pt]
\hline
$f_{0.3-8}$ & \ \ \ \ $\left(5.1^{+0.8}_{-0.8}\right) \times 10^{-15}$ erg cm$^{-2}$ s$^{-1}$\\[3pt]
$L_{0.3-8}$ & \ \ \ \ $\left(9.3^{+1.5}_{-1.5}\right) \times 10^{36}$ erg s$^{-1}$\\[3pt]
EM & $\left(4.5^{+3.5}_{-1.9}\right) \times 10^{60}$ cm$^{-3}$\\[3pt]
\hline
C-statistic &  \ \ \ \  $9.77$ (17 dof)\\[3pt]
\hline
\end{tabular}
\end{center}
\caption{Best-fitting spectral parameters for the X-ray emission 
from the cocoon (not including the hot spots).
The XSPEC models are {\it wabs$_{\rm Gal}$*ray} and {\it wabs$_{\rm Gal}$*brems}.
Adding intrinsic absorption does not improve the fit.
Errors are 90\% confidence level for 1 interesting parameter.}
\label{tab:a}
\end{table}

\section{Discussion}

\subsection{Energetics of the bubble}

We have presented radio and X-ray results from our multiband
study of a powerful non-nuclear BH in NGC\,7793, and
of its surrounding shock-ionized cocoon (see PSM10 
for a discussion of the evidence for shock ionization 
from the optical emission lines). 
The system was originally classified as a supernova remnant \citep{bla97}.
In that scenario, \citet{asv06} showed
that an input energy $\approx 5 \times 10^{52}$ erg was required
to explain its size and radio luminosity, well beyond
the energy that can be supplied by an individual supernova.
Based on the clear radio and X-ray evidence 
for a collimated jet pair (lobes, hot spots), PSM10 showed 
that such a large amount of energy has been supplied 
by the BH over the lifetime of the bubble 
(characteristic age $\approx 2 \times 10^5$ yr).
The core is seen as a faint point-like X-ray source, consistent
with a stellar-mass BH in the low/hard state,
and a point-like optical source, consistent
with an OB donor star (possibly a Wolf-Rayet: PSM10). It is undetected
in the radio bands, to a $3\sigma$ upper limit $\approx 0.03$ mJy.
This is unsurprising: if the BH lies in the fundamental plane 
\citep{mer03,kor06} with a mass $\sim 10 M_{\odot}$, we expect
a core radio flux $\sim 0.01 \mu$Jy ($\nu L_\nu \sim 10^{30}$ erg s$^{-1}$),
like from a common-or-garden 
low/hard state microquasar at a distance of 3.9 Mpc.
On the other hand, the impact of this BH onto
the surrounding interstellar medium is all but common.
The simplest explanation is that the core is currently
in a low/hard state, three or four orders of magnitude fainter
than its long-term average power. In that same canonical state, 
the steady jet power ${\mathcal P} \propto L_{\rm X}^{0.5}$ 
\citep{fen04,mal04,fen03}. The normalization of this relation 
has an uncertainty of almost two orders of magnitude, but is 
constrained enough to suggest $10^{36}$ erg s$^{-1}$ 
$\la {\mathcal P} \la 10^{38}$ erg s$^{-1}$, also much 
lower than the inferred long-term average. However, there
may be alternative explanations for the apparent faintness 
of the core. Perhaps the observed X-ray luminosity is a severe underestimate 
of the true X-ray luminosity, if most of the direct emission is absorbed
and/or beamed away from our line of sight \citep[as it has been suggested 
for SS\,433:][]{med10} and we are only seeing a scattered component. Or perhaps
the system is not in the canonical low/hard state, but
in some other unclassified state with ${\mathcal P} \gg L_X$.

The faint core is in stark contrast with the large, bright
nebula, visible in all bands with a similar size and shape
(Figure 5), and a  conservatively estimated volume $\approx 10^{62}$ cm$^{-3}$, 
assuming a prolate spheroid with a major axis $\approx 280$ pc 
and minor axes $\approx 130$ pc (based on the projected distance between 
the hot spots along the major axis, and the width of the radio nebula 
at 5.5 GHz in the transverse direction). In fact, H$\alpha$ images (PSM10) 
may suggest an even larger size, $\approx 340 \times 170$ pc.
Thus, the volume-averaged shell radius $R_{\rm s} \approx 100$ pc.
Its characteristic size is an order of magnitude larger than
the jet driven bubble around Cyg X-1, which has an estimated
jet power $\sim 10^{37}$ erg s$^{-1}$ \citep{rus07}.
It is a factor of two larger (allowing for distance uncertainties)
than the radius of the SS433/W50 nebula, with an estimated jet power
$\sim 10^{39}$ erg s$^{-1}$ \citep{med10,fab04,mar02}.

PSM10 determined a mechanical power ${\mathcal P} \approx 5 \times 10^{40}$ 
erg s$^{-1}$ for S26, using  
the well-known self-similar solution to the conservation 
of mass, momentum and energy equations \citep[Equations 17--22 in][]{wea77}, 
in which the radius of the swept-up shell 
$R_{\rm s} \approx 0.76 \,{\mathcal P}^{1/5}t^{3/5}\rho_{\rm 0}^{-1/5}$.
They measured the expansion velocity 
from the half-width at zero-intensity of the optical emission lines 
($v \approx 250$ km s$^{-1}$) and from shock-ionization models of the  
He{\footnotesize{II}} $\lambda 4686$/H$\beta$ 
flux ratio ($v \approx 275$ km s$^{-1}$). This implies 
a characteristic age $t = (3R_{\rm s})/(5v_{\rm exp}) \approx 2 \times 10^5$ yr.
The hydrogen number density of the interstellar medium into which the bubble 
expands was estimated as $n_0 \approx 0.7$ cm$^{-3}$ (PSM10), 
from a comparison of the observed H$\alpha$ emission with 
the intensity of a fully radiative shock \citep{dop96,pak06}; 
this corresponds to a mass density $\rho_{\rm 0} \approx \mu m_{\rm p}n_0 
\approx 1.6 \times 10^{-24}$ g cm$^{-3}$ (taking the mean atomic 
weight $\mu = 1.38$).
The swept-up mass in the expanding shell is $(4\pi/3)\rho R_{\rm s}^3 
\approx 2 \times 10^{38}$ g, carrying a kinetic energy 
$\simeq (15/77) {\mathcal P} t \approx 6 \times 10^{52}$ erg.
The energy content 
of the thermal gas between the reverse shock and the swept-up shell 
is $E = (5/11) {\mathcal P} t \approx 10^{53}$ erg.


\begin{figure}
\psfig{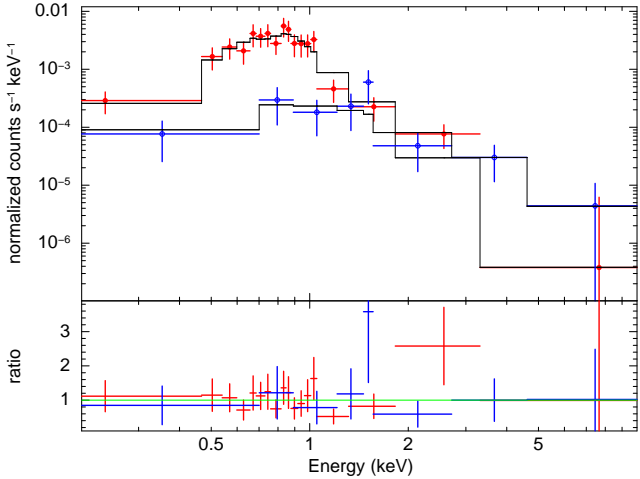}
\caption{{\it Chandra}/ACIS spectra of the combined hot spot emission 
(red datapoints) and core emission (blue datapoints), fitted 
with a single-temperature {\it sedov} thermal-plasma model 
and with a power-law model, respectively.
See Tables 2, 3, 4 for the best-fitting parameters. 
The {\it sedov} model fit illustrated here is statistically equivalent 
to the two-temperature {\it raymond-smith} model fit 
illustrated in Fig.~2 of PSM10.
}
\label{f1}
\end{figure}

\begin{figure}
\psfig{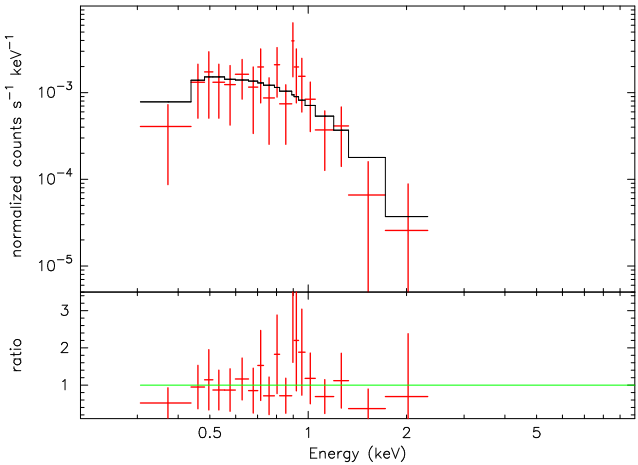}
\caption{{\it Chandra}/ACIS spectrum of the cocoon emission, 
fitted with a $\approx 0.5$ keV bremsstrahlung model. 
See Table 5 for the best-fitting parameters.}
\label{f1}
\end{figure}



The only shock-ionized nebulae of comparable size and energy content
in the local universe are those around ULXs such as 
Holmberg IX X-1, NGC\,1313 X-2, IC\,342 X-1 \citep{pak02,pak08,ram06,fen08}.
One difference is that, unlike all previously known ULX bubbles,
S26 has clear evidence of collimated jets. We do not know 
the relative distribution of mechanical power between the collimated jets 
and perhaps a more spherically symmetric wind (for example an accretion 
disk wind); however, the elongated structure of the nebula and the presence 
of bright lobes and hot spots suggests that the jet carries most of the power.
Another difference is that in ULX bubbles, the central BH
is (by definition) X-ray luminous, with an apparent X-ray luminosity
$\sim 10^{40}$ erg s$^{-1}$, similar to the mechanical power.
S26 may be an example of ULX bubble where the central BH
is currently in a low and/or jet-dominated state.
In that respect, S26 is analogous to (but two orders of magnitude 
more energetic than) the shock-ionized bubble 
around the very massive but only moderately
luminous non-nuclear BH IC\,10 X-1 \citep{pre07}.

\subsection{Synchrotron and thermal plasma emission}

The detection of a radio cocoon with bright radio hot spots
shows that some of the input mechanical power goes into
synchrotron-emitting, relativistic electrons.
As expected, the radio spectrum is steep in the lobes
and flat or inverted near the position of the core (Figure 3); 
this may be evidence of recent or continuing ejection activity 
(although the core itself is undetected in the radio).
In addition, there is radio emission from the cocoon region 
outside the lobes, with a specific flux $\approx 0.7$ mJy at 5.5 GHz.
There is circumstatial evidence that this emission 
has a rather flat spectrum, certainly 
flatter than in the lobes (Figure 3). This is difficult 
to reconcile with a scenario where the synchrotron-emitting 
electrons in the whole cocoon are backflowing from 
the lobes. In that case, the spectral index in the rest 
of the cocoon would be even steeper---as we see for example 
in Cygnus A \citep{car96}. 
We suggest that the extended radio emission in the cocoon 
outside the lobes may have a significant contribution 
from (flat-spectrum) free-free emission, 
from the same thermal gas responsible for the optical 
recombination lines. 
For the characteristic temperature $\approx 3 \times 10^4$ K 
estimated by PSM10, the ratio of the H$\beta$ and free-free 
radio emissivity is $j_{H\beta}/j_{5.5 \rm GHz} \approx 2 \times 10^{-10}$ 
erg cm$^{-2}$ s$^{-1}$ Jy$^{-1}$ \citep[][their Appendix A]{cap86}.
The H$\beta$ luminosity of S26 is $\approx 10^{38}$ erg s$^{-1}$ 
(PSM10): thus, we expect a free-free radio flux $\approx 0.3$ mJy 
at 5.5 GHz. This is negligible in the lobes, compared with 
the synchrotron component, but may be significant 
in the region outside the lobes, and may explain 
the rather flat spectral index there.

What fraction of the mechanical power is transferred
to relativistic electrons? If we combine the self-similar
model of cocoon expansion with the standard synchrotron 
emissivity, in the minimum-energy approximation, 
assuming a spectral index $\alpha = -0.7$, we obtain (Appendix A):
\begin{equation}
S_{\nu} \approx 82 \, (1+k)^{-1} \, \eta^{1.85} \, {\mathcal P_{39}}^{1.34} \, 
       t_5^{0.32} \, n_1^{0.51} \, d_1^{-2} \, \nu_5^{-0.7} 
      \, {\rm mJy},
\end{equation}
where 
$\eta$ is the fraction of the total energy density contained in all 
relativistic species (electrons, protons and nuclei) plus magnetic field, 
$(1+k)^{-1}$ is the fraction of relativistic particle energy  
carried by the synchrotron-emitting electrons alone,
${\mathcal P_{39}}$ is the jet power in units of $10^{39}$ erg s$^{-1}$, 
$t_5$ is the source age in units of $10^5$ yr, $n_1$ is the interstellar 
number density in cm$^{-3}$, $d$ is the source distance in Mpc, 
and $\nu_5$ the observed frequency in units of 5 GHz. 
A specific flux a few times higher is expected if we assume 
$\alpha = -0.5$ (Appendix A).

If we assume that all the jet power is transferred to the relativistic 
electrons (that is, if we put $k = 0$ and $\eta = 1$), 
Equation (1) grossly overestimates the radio emission, 
for the measured jet power and distance of S26. (Or, conversely, 
the observed radio flux would lead us to underestimate the jet power 
if we did not know it independently).
This tells us that $(1+k)^{-1} \times \eta^{1.85} \sim 10^{-3}$.
We cannot separately determine $k$ and $\eta$ from this simple model, 
but for plausible values of $k \sim 10$--$100$ found in cosmic rays, 
we estimate that the fraction $(1+k)^{-1} \times \eta$ 
of the total injected mechanical power carried by the relativistic electrons
is $\sim$ a few $10^{-3}$.
The rest of the energy is given to protons, nuclei and non-relativistic electrons,  
and is used for heating and inflating the bubble, 
accelerating the shell of swept-up interstellar medium 
to the expansion speed $\approx 250$ km s$^{-1}$.


The X-ray emission from the hot spots and cocoon provides another 
clue to understand the energy budget.
In radio galaxies, X-ray hot spots are usually interpreted 
either as direct synchrotron, or synchrotron self-Compton emission 
\citep{har02,har04}, from the same population of electrons responsible 
for the radio hot spots, which are accelerated at the reverse shock.
For S26, we estimate from {\it Chandra} a specific flux 
$\approx 10^{-14}$ erg cm$^{-2}$ s$^{-2}$ keV$^{-1}$ at 1 keV from both hot spots, 
corresponding to $\approx 4 \times 10^{-32}$ erg cm$^{-2}$ s$^{-2}$ Hz$^{-1}$.
The combined radio emission from the radio hot spots 
is $\approx 1$ mJy $\approx 10^{-26}$ erg cm$^{-2}$ s$^{-2}$ Hz$^{-1}$ at 5.5 GHz.
If we extrapolate the radio flux as a straight power law, 
$S_\nu \sim \nu^{-0.7}$, 
we would also expect $\approx 4 \times 10^{-32}$ erg cm$^{-2}$ s$^{-2}$ Hz$^{-1}$ 
at 1 keV. This of course requires a continuous acceleration 
of the most energetic electrons, so that there is no spectral break 
from the radio to the X-ray bands.
So, in principle, the X-ray to radio flux ratio is consistent 
with a simple synchrotron component.
However, the X-ray spectrum tells a different story. Its shape 
and slope are not consistent 
with either synchrotron or inverse-Compton power-law models, 
even accounting for the low number of counts. We showed (Section 4 and Figure 6) 
that the X-ray emission from the hot spots, with its peak 
at $\sim 0.6$--$0.9$ keV and its sharp drop above $\approx 1$ keV, 
is most likely due to hot thermal plasma with a range of temperatures 
up to $\approx 0.9$ keV.
We also showed that the X-ray hot spots are located $\approx 1\arcsec$ further 
away from the core than the radio hot spots.
This is a second argument in support of our claim that X-ray and radio 
hot spots are due to different physical processes.

\begin{figure}
\psfig{figure=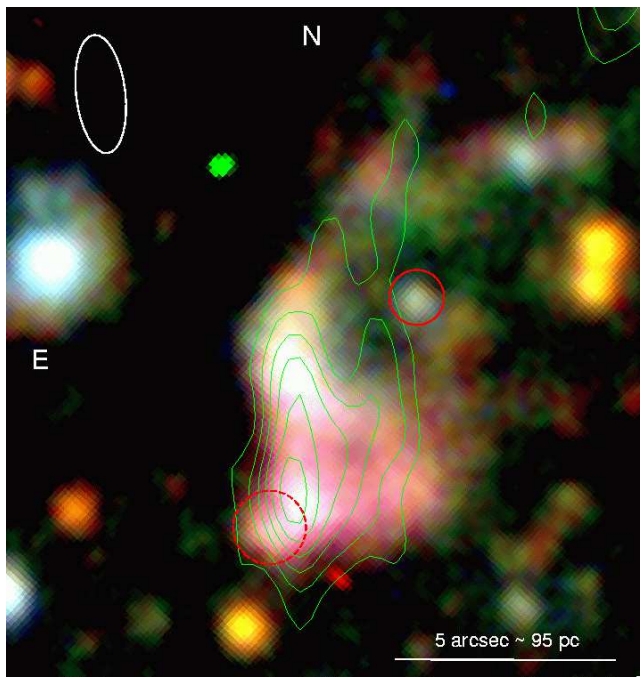,width=84mm,angle=-0}
\caption{Close-up view of the southern lobe: true-colour image in the $B,V,R$ bands, 
taken by Dr.~Jifeng Liu with the Baade {\it Magellan} telescope on 2009 August 28 
(exposure time: 200 s per filter). The ATCA 9.02-GHz intensity contours 
are overplotted in green; the red circles mark the position of the X-ray core 
and southern hot spot. The point-like optical core has a brightness $B \approx 23$ mag, 
$M_B \approx -5$ mag (PSM10).}
\label{f1}
\end{figure}

It is still not clear what is heating at least part of the X-ray 
emitting gas to such high temperatures, particularly 
at the hot spots. We suggest two alternative scenarios.
The first scenario is that the X-ray emitting gas is the shocked 
interstellar medium, heated by the bow shock 
(advancing at a speed $v_{\rm bs}$) 
to a temperature $kT = (3/16) \mu m_{\rm p} v_{\rm bs}^2$. 
In this case, the X-ray hot spots mark the position of the bow shock and 
the radio hot spots that of the reverse shock into the ejecta.
To produce temperatures $\approx 0.9$ keV (as required by our fits with 
{\it {raymond-smith}} and most other thermal plasma models 
in {\footnotesize XSPEC}), the bow shock velocity 
(that is, the expansion velocity along the major axis) 
would have to be $\approx 900$ km s$^{-1}$, almost 4 times 
higher than the expansion velocity measured by PSM10 
from the width of the optical lines. However, their slit position 
was almost parallel to the minor axis and did not include the hot spots; 
moreover, the viewing angle of the major axis is still unknown. 
Thus, we still do not know at what speed the jet heads are advancing 
into the interstellar medium,  Besides, we noted in Section 4.1 that 
a {\it {sedov}} model gives a good fit of the hot spot emission 
with thermal plasma temperatures as low as $\approx 0.5$ keV, 
requiring a more plausible shock velocity $\approx 300$ km s$^{-1}$.
To sum up, we cannot yet rule out the fast bow shock scenario for S26.
An example of an X-ray-emitting bow shock located 
ahead of the radio-emitting lobes can be seen in the nearest radio galaxy, 
Cen A \citep{kra07}.
An alternative scenario, considered more likely by PSM10, 
is that the bow shock is not advancing 
fast enough to produce the X-ray emitting gas, and the shocked interstellar gas 
between the bow shock and the contact discontinuity has already 
cooled and collapsed to a thin, dense shell. In this case, 
the X-ray emitting gas is located between the reverse shock 
and the swept-up outer shell,  
and is heated by the shocked ejecta via thermal conduction. 
Most of the mass in the hot region must come from 
mass-loading of denser interstellar clouds during the bubble expansion, 
and from the evaporation of part of the swept-up shell of interstellar 
medium, and its mixing with the lower-density, hotter jet material.

The physical size of the X-ray hot spots is $\la 20$ pc in radius, 
and the projected size of the whole bubble is $\approx 300 \times 150$ pc.
From the estimated emission measures (Tables 3--5), 
we infer a mass of X-ray emitting gas $\sim$ a few $100 M_{\odot}$ 
in the hot spots (see also PSM10) and $\sim$ a few $1000 M_{\odot}$ in the cocoon 
(assuming a filling factor $\sim 1$). These values are several 
orders of magnitude higher than the mass that could have been 
carried out by the BH jet and winds over the source lifetime. 
But the mass of X-ray emitting gas is an order of magnitude 
less than the total mass of the swept-up interstellar medium; 
that is, the swept-up shell is not significantly depleted 
by evaporation into the hot region, in agreement 
with the self-similar approximation of \citet{wea77}.
From the estimated mass and fitted X-ray temperatures, 
we conclude that the X-ray-emitting gas contains 
a thermal energy $\sim 10^{51}$ erg (hot spots) 
and $\sim 10^{52}$ erg (whole cocoon). And we showed earlier that 
the total thermal energy $\sim 10^{53}$ erg and the energy 
carried by the synchrotron-emitting relativistic 
electrons $\sim$ a few $10^{50}$ erg.
The cooling timescale of the X-ray emitting gas 
in the cocoon is $\sim 10^7$ yr, 100 times longer than the age of the source. 
This is consistent with an emitted X-ray luminosity $\sim 10^{37}$ erg s$^{-1}$ 
even though power may have been transferred from the jet 
to the $\sim 0.3$--$1$ keV component 
of the gas at an average rate $\sim 10^{39}$ erg s$^{-1}$.  
By analogy with other shock-heated bubbles, we suggest  
that there may be even hotter but much less dense gas 
components, especially near the hot spots, whose hard X-ray emission 
emission would be too faint to be detectable in the 50-ks {\it Chandra} 
observation.

\section{Conclusions}

We have carried out a multiband study of the shock-ionized 
bubble S26 in NGC\,7793, which looks like a long-sought analogue 
of the Galactic jet source SS\,433/W50 but on an even grander scale 
(projected size $\approx 300 \times 150$ pc). 
We showed that its structure is a scaled-down 
version of powerful FRII radio galaxies, with a core, radio lobes, 
X-ray hot spots and cocoon. It is the first time that all these elements 
have been found in a non-nuclear BH. 
We showed that the radio and X-ray hot spots are not spatially 
coincident: the X-ray hot spots are $\approx 20$ pc further out 
than the peak of the radio intensity in the lobes. 
This suggests that X-ray and radio emission come from 
different populations of radiating particles.
Based on our {\it Chandra} spectral analysis, we argued that the X-ray emission 
from the hot spots is most likely thermal. From the ATCA data, 
we showed that the radio emission from the lobes 
has a steep spectrum, consistent with optically-thin synchrotron 
emission. Over the rest of cocoon, the radio spectrum is flatter, 
suggesting an additional contribution from free-free emission; this is 
consistent with what we would expect from the measured H$\beta$ 
line emission. A point-like radio core is not detected, but the radio 
spectrum is flat or inverted in the proximity of the X-ray/optical core 
position; this may be interpreted as a more recent ejection. However, 
deeper ATCA observations are needed (and scheduled) to test 
this suggestion.

The total particle energy in the bubble is $\sim 10^{53}$ erg. 
Based on the measured radio flux and size of the bubble, 
and using standard equipartition relations for microquasar lobes, 
we estimated that the energy carried by the synchrotron-emitting 
relativistic electrons is a few 100 times less 
than the energy stored in protons, nuclei and non-relativistic electrons; 
non-relativistic particles provide most of the pressure to inflate the bubble. 
This system can give us important clues on how BHs 
at near-Eddington accretion rates transfer energy 
to the surrounding medium. The size and total energy content  
of the bubble are comparable to those found in some ULXs.
However, here the core appears to be currently X-ray faint (and was so 
also during the {\it Einstein} and {\it ROSAT} observations), while 
the jet is carrying a long-term-average power $\sim 10^{40}$ erg s$^{-1}$ (PSM10).
We do not have any information on the long-term-average X-ray luminosity 
of the core, so we cannot exclude that it is similar to the mechanical power. 
If the BH in S26 is of stellar origin, its super-Eddington jet power 
may force us to rethink the ``canonical'' scheme of BH accretion states. 
In Galactic BH transients, a collimated jet is present 
at accretion rates $\la$ a few percent of the Eddington rate (low/hard state). 
At higher accretion rates ($\sim 0.05$--$0.5$ Eddington), the accretion flow 
usually collapses to a geometrically-thin, 
radiatively-efficient thermal disk, and the jet is quenched.
At even higher accretion rates (above Eddington), high X-ray luminosity 
and powerful mass-loaded outflows may coexist, but it is not known whether 
there can also be steady, collimated jets, and what their power is compared 
with the radiative power. S26 suggests that there can be collimated jets, 
and they may even dominate over the radiative output.
The same scenario has been suggested for some powerful FRII radio galaxies 
and quasars \citep{pun07,ito08}.

%

\section*{Acknowledgments}
We thank: Tasso Tzioumis for his assistance when we prepared the ATCA observations; 
Jifeng Liu for his taking of an optical image for us, using his own time 
at the Magellan telescope; 
Mike Dopita for comments and for his taking of an optical spectrum for us, 
from his own time at the ANU 2.3m telescope; Geoff Bicknell, Fabien Gris\'{e}, JingFang Hao, 
Albert Kong, Zdenka Kuncic and Kinwah Wu for comments.
RS acknowledges hospitality at the National Tsing Hua University (Taiwan), 
at the Institute for High Energy Physics (Beijing), and at the University of Sydney, 
during part of this work. The Australia Telescope is funded by the  
Commonwealth of Australia for operation as a National Facility managed  
by CSIRO.


\appendix

\section[]{Synchrotron emission in the minimum-energy condition}

To estimate the minimum energy associated with the synchrotron-emitting 
cocoon, we assume an energy range 
$(\gamma_{\rm min},\gamma_{\rm max})$ for the relativistic 
electrons \citep{poh93,bic05}, rather than a frequency range. 
Typical empirical values of $\gamma_{\rm min} \sim 1$--$10$ \citep{blu00}
and $\gamma_{\rm max} \sim 10^4$--$10^5$.
For a steep spectrum, the minimum energy depends only 
very weakly on the high-energy cut-off.
We introduce the following quantities:  
$\epsilon_e$ is the energy density in relativistic electrons; 
$\epsilon_p \equiv (1+k) \epsilon_e$ is the energy density in relativistic 
particles (electrons and protons), where $k$ is a free parameter; 
$\epsilon_B$ is the energy density in the magnetic field;
$\epsilon_{\rm tot} = \epsilon_p + \epsilon_B$  
includes the energy in relativistic particles and magnetic field; 
$\epsilon'_{\rm tot}$ is the total energy density including  
relativistic and non relativistic particles and the field.
We introduce another free parameter $\eta \equiv \epsilon_{\rm tot}/\epsilon'_{\rm tot}$ to express 
the relative fraction of total energy stored in relativistic particles 
(protons and electrons) plus field. 
The main reason why we distinguish between $\epsilon_{\rm tot}$ and $\epsilon'_{\rm tot}$
is that there is solid observational evidence \citep{cav10,pun07,lea01,wil99} 
that most of the energy in the lobes and cavities of radio galaxies 
is in low-energy electrons and other non-relativistic particles 
({\it i.e.} $\eta  \ll 1$), and that the energy density of the magnetic 
field may be $\sim 10$--$100$ times less than the total energy density $\epsilon'_{\rm tot}$.
Thus, using $\epsilon'_{\rm tot}$ to derive a minimum-energy or equipartition 
criterion generally leads to very inaccurate estimates for the synchrotron 
emission of a lobe. Instead, here we apply those criteria only 
to $\epsilon_{\rm tot}$, that is we assume that magnetic energy density 
is of the same order of magnitude as the relativistic particle 
energy density but much less than the total particle energy density. 
And within the relativistic energy density component, we use the parameter 
$k$ to express the relative contribution of nuclei and electrons.

Applying the minimum-energy condition leads, after some algebra \citep{bic05}, to this 
expression for the minimum-energy magnetic field:
\begin{eqnarray}
B^2_{\rm min} &=&  \left(\frac{m_e}{e}\right)^2 \left[ \frac{p+1}{2} \, (1+k) \,
	C^{-1}(p) \, \frac{c}{m_e} \right]^{4/(p+5)} 
	\nonumber\\
	&\times&
	\left[h(p,\gamma_{\rm min}, \gamma_{\rm max})\,  
	\frac{I_{\nu} \nu^{(p-1)/2}}{2r_{\rm c}}\right]^{4/(p+5)}
	\nonumber\\
	&\approx& \left(\frac{m_e}{e}\right)^2 \, \left(\frac{c}{m_e} \right)^{4/(p+5)} \,
	\left(\frac{3}{4\pi} \right)^{4/(p+5)}	\nonumber\\
	&\times&
	\left[ \frac{p+1}{2} \, (1+k) \, C^{-1}\,h\right]^{4/(p+5)} \, 
	 d^{8/(p+5)} \, r_{\rm c}^{-12/(p+5)} 
	\nonumber\\
	&\times&
	S_{\nu}^{4/(p+5)} \, \nu^{2(p-1)/(p+5)}
\end{eqnarray}
where the energy spectrum of the electrons is $N(E)dE \sim E^{-p}dE$, 
$m_e$ and $e$ are the electron mass and charge, $c$ the speed of light, 
$I_{\nu}$ the specific surface brightness 
$S_{\nu}$ the specific flux at the observer's position, integrated over the whole 
cocoon, $r_{\rm c}$ is the cocoon radius, and $d$ is the distance to the source.
We have assumed a filling factor of 1, for simplicity.
The functions 
\begin{equation}
h(p,\gamma_{\rm min}, \gamma_{\rm max}) = \frac{1}{p-2}\,
	\left[\gamma_{\rm min}^{(2-p)} - \gamma_{\rm max}^{(2-p)}
	\right],
\end{equation}
\begin{eqnarray}
C(p) &=& \frac{3^{p/2}}{2^{(p+13)/2}\,\pi^{(p+2)/2}} \nonumber\\
	&\times& \frac{\Gamma\left(\frac{p}{4}+\frac{19}{12}\right) \,
	\Gamma\left(\frac{p}{4}-\frac{1}{12}\right) \,
	\Gamma\left(\frac{p}{4}+\frac{1}{4}\right)}
	{\Gamma\left(\frac{p}{4}+\frac{7}{4}\right)},	
\end{eqnarray}
and $\Gamma(z)$ is the Gamma function.
The corresponding total (minimum) energy density is:
\begin{equation}
\epsilon_{\rm tot, min} 
        = \left[(1+k) \epsilon_e + \epsilon_B\right]_{\rm min} =
	\left(\frac{4}{p+1} + 1\right)
	\left(\frac{B_{\rm min}^2}{2\mu_0}\right)
\end{equation}

%

We now need to relate the energy density $\epsilon_{\rm tot, min}$
to the input jet power and size of the bubble. 
An approximate expression we could use is that the total energy 
(relativistic, non relativistic and field) is simply ${\mathcal P} t$.
However, this is not entirely correct, because part of the injected 
energy is spent to inflate the bubble. From the self-similar solution 
of \citet{wea77}, we obtain a more accurate expression 
for the energy still available:
\begin{equation}
\epsilon'_{\rm tot} \equiv \frac{3}{4\pi} \frac{5}{11} {\mathcal P} t r_{\rm c}^{-3}, 
\end{equation}
and according to our definition of $\eta$, 
\begin{equation}
\epsilon_{\rm tot} \equiv \frac{3}{4\pi} \frac{5}{11} \eta {\mathcal P} t r_{\rm c}^{-3}. 
\end{equation}
In the minimum-energy approximation, from Eq.~A4:
\begin{equation}
\left(\frac{B_{\rm min}^2}{2\mu_0}\right) = \frac{3}{4\pi} \frac{5}{11} \eta 
   \left(\frac{p+1}{p+5}\right) {\mathcal P} t r_{\rm c}^{-3},
\end{equation}
and this value can now be substituted into Eq.~A1.
Finally, the cocoon radius $r_{\rm c}$ is obtained 
from the \citet{wea77}'s set of self-similar solutions 
(assuming a thin outer shell):
\begin{equation}
r_{\rm c} \simeq \left(\frac{125}{154\pi}\right)^{1/5}
\times \left(\frac{{\mathcal P}t^3}{\rho_{\rm 0}}\right)^{1/5}
\simeq
0.76 \times \left(\frac{{\mathcal P}t^3}{\rho_{\rm 0}}\right)^{1/5},
\end{equation}
and this expression is also substituted into Eq.~A1.
(Note that here it is the total jet energy ${\mathcal P} t$ 
that determines the size of the bubble).

Rearranging Eq.~A1 with such substitutions, we obtain:
\begin{eqnarray}
S_{\nu} &\approx& 1.84 \, \left(0.40 \, \eta \, \frac{p+1}{p+5}\right)^{(p+5)/4} \,
                  \left(\frac{2\mu_0e^2}{m_e^2}\right)^{(p+5)/4} \frac{m_e}{c} \nonumber\\
	&\times& \left[ \frac{p+1}{2} \, (1+k) \, C^{-1}\, 
		    h(p,\gamma_{\rm min}, \gamma_{\rm max})\right]^{-1}  \nonumber\\
	&\times&   {\mathcal P}^{(p+11)/10} \, t^{(4-p)/5}  \,
	  \rho^{3(p+1)/20} \, \nu^{(1-p)/2} \, d^{-2},
\end{eqnarray}
where the numerical values of $h$ and $C$ come from Eqs.~(A2,3). 
For a spectral index $\alpha \approx -0.5$ ($p \approx 2$), $\gamma_{\rm min} \la 10$ 
and $\gamma_{\rm max} \sim 10^5$, we have, in physical units:
\begin{equation}
S_{\nu} \approx 640 \, (1+k)^{-1} \, \eta^{7/4} \, {\mathcal P_{39}}^{1.3} \, 
       t_5^{0.4} \, n_1^{0.45} \, d_1^{-2} \, \nu_5^{-0.5} 
      \ \ {\rm mJy},
\end{equation}
where 
${\mathcal P_{39}}$ is the jet power in units of $10^{39}$ erg s$^{-1}$, 
$t_5$ is the source age in units of $10^5$ yr, $n_1$ is the interstellar 
number density in cm$^{-3}$, $d$ is the source distance in Mpc, 
$\nu_5$ the observed frequency in units of 5 GHz; the numerical 
coefficient is not very sensitive to the choice of $\gamma_{\rm min}$.
We can obtain an analogous estimate for a spectral index $\alpha \approx -0.7$  
(corresponding to $p \approx 2.4$), which is more often the case in 
radio lobes. In that case, 
\begin{equation}
S_{\nu} \approx 82 \, (1+k)^{-1} \, \eta^{1.85} \, {\mathcal P_{39}}^{1.34} \, 
       t_5^{0.32} \, n_1^{0.51} \, d_1^{-2} \, \nu_5^{-0.7} 
      \, {\rm mJy},
\end{equation}
where we have fixed this time $\gamma_{\rm min} = 1$.

We can now compare these specific fluxes with the observations: 
S26 has a 5.5-GHz flux $\approx 2$ mJy, for a jet power 
$\sim$ a few $10^{40}$ erg s$^{-1}$, at a distance of $3.9$ Mpc.
This tells us that $(1+k)^{-1} \times \eta^{1.85} \sim 10^{-3}$, 
that is the energy stored in synchrotron-emitting relativistic electrons 
(a fraction $\eta/(1+k)$ of the total) is much less than the energy stored 
in relativistic protons and in non-relativistic particles.  
We cannot determine the individual values of $\eta$ and $k$ from 
this set of radio observations alone: only their combination. 
If we use cosmic rays 
as an analogy (since they may be accelerated in jet and supernova shocks), 
we would expect $k \sim 100$. For plausible values $k \sim 10$--$100$, 
$\eta/(1+k) \sim$ few $10^{-3}$.

\end{document}